\documentstyle[manuscript,prl,aps]{revtex}

\begin{document}

\title{\bf Dimensionality dependence of optical nonlinearity and relaxation
dynamics in cuprates}

\author{M. Ashida,$^1$ Y. Taguchi,$^2$ Y. Tokura,$^{2,3}$ 
R.T. Clay,$^{1,4}$ S. Mazumdar,$^4$ Yu. P. Svirko,$^1$ 
and M. Kuwata-Gonokami$^{1,2}$\cite{auth}}
\address{$^{1}$Cooperative excitation Project, ERATO, Japan Science and \\
Technology Corporation (JST), Kanagawa 213-0012, Japan}
\address{$^{2}$Department of Applied Physics, the University of Tokyo, Tokyo
113-8656, Japan}
\address{$^{3}$Correlated Electron Research Center (CERC) and Joint 
Research Center for Atom Technology (JRCAT), Tsukuba 305-8562, Japan}
\address{$^{4}$Department of Physics and the Optical Sciences Center, 
University of Arizona, Tucson, AZ 85721}

\date{\today}
\maketitle

\begin{abstract}
Femtosecond pump-probe measurements find pronounced dimensionality 
dependence of the optical nonlinearity in cuprates. Although the 
coherent two-photon absorption (TPA) and linear absorption bands 
nearly overlap in both quasi-one and two-dimensional (1D and 2D) 
cuprates, the TPA coefficient is one order of magnitude smaller in 
2D than in 1D. Furthermore, picosecond recovery of optical 
transparency is observed in 1D cuprates, while the recovery 
in 2D involves relaxation channels with a time scales of tens of 
picoseconds. The experimental results are interpreted within the 
two-band extended Hubbard model.
\end{abstract}

Transport and magnetic behavior of strongly correlated electron (SCE) systems, 
which depend on the lowest excitations, have been studied extensively in recent
years. High energy electronic excitations, in contrast, have received less 
attention. Linear absorption studies, performed for the copper oxide based SCE 
systems \cite{Tokura}, do not reveal features associated with optically dark 
two-photon states. Femtosecond spectroscopic studies in quasi-two dimensional 
(2D) cuprates \cite{Matsuda,Kaindl} have revealed the strong role played by 
magnetic excitations in ultrafast non-radiative relaxation processes, which is
related to the large Heisenberg exchanges in the cuprates 
\cite{Perkins,Lorenzana,Motoyama,Suzuura,Imada}. More recently, interest has 
shifted to nonlinear optical studies which provide information about both odd 
parity one-photon states and even parity two-photon states 
\cite{Ogasawara,Kishida,Schulzgen}.

An additional motivation to study nonlinear optical properties of SCE materials
originates from the continued quest for new materials with ultrafast and strong
optical nonlinearity. The mechanism of optical nonlinearity in SCE systems 
\cite{Dixit} is different from that in band insulators, and it is conceivable 
that new SCE materials can be found with nonlinear optical properties that 
meet the requirements for technological applications. Initial support for this 
idea has come from the demonstrations of strong two-photon absorption (TPA) 
along with picosecond recovery of optical transparency in the quasi-one 
dimensional (1D) cuprate Sr$_2$CuO$_3$ \cite{Ogasawara}, and of giant 
electro-reflectance in several 1D Mott insulators \cite{Kishida}. An important 
new question that has arisen involves the role of dimensionality in SCE 
nonlinear optical materials. In conventional band insulators confinement of the
Wannier exciton in 1D can strongly enhance the optical nonlinearity 
\cite{Chemla}. Whether or not similar dimensionality effects occur in SCE 
systems has not been investigated systematically so far. In the present work we
compare the results of the sub-picosecond pump-probe transmission measurements 
in 1D and 2D cuprates. We observe significant dimensionality dependences in the
magnitude of the nonlinear response as well as the relaxation dynamics, which 
are explained using cluster calculations within the appropriate extended 
Hubbard Hamiltonian.

To clarify the dimensionality dependence systematically, we selected three
materials, which are well-known strongly correlated 1D and 2D charge-transfer 
(CT) insulators. Specifically, Sr$_2$CuO$_3$ (see Fig.~1(a)) and SrCuO$_2$ 
(Fig.~1(b)) have single and weakly coupled double Cu-O chains, respectively 
\cite{Motoyama,Imada}. Since in SrCuO$_2$ the interchain coupling due to 
90$^\circ$ Cu-O-Cu bonds is much weaker than the coupling along the 180$^\circ$
intrachain Cu-O-Cu bonds, this material can be also classified as a 1D spin 
system \cite{Motoyama,Imada}. Sr$_2$CuO$_2$Cl$_2$ (Fig.~1(c)) is a 2D 
cuprate with the same CuO$_2$ network as in La$_2$CuO$_4$ with the apical 
oxygens of the latter replaced with Cl ions \cite{Tokura}. Importantly, the 
lattice constants in the Cu-O chain direction in these materials are almost 
same. All of these materials show optical gaps around 2 eV, which corresponds 
to CT excitation, and are nearly transparent below 1.5 eV (see Fig.~3). The 
strong antiferromagnetic exchange interaction between holes on neighboring
Cu sites, $J \sim$ 2000 - 3000 K for Sr$_2$CuO$_3$ and SrCuO$_2$ 
\cite{Motoyama,Suzuura,Imada} and $J \sim$ 1400 K for Sr$_2$CuO$_2$Cl$_2$ 
\cite{Perkins,Lorenzana}, gives rise to wide spinon and magnon bands in 1D 
and 2D materials, respectively. In spite of the one-dimensionality of 
Sr$_2$CuO$_3$, a spin-Peierls transition does not occur in this system.

Single crystals of Sr$_2$CuO$_3$ and SrCuO$_2$ are grown by the traveling-
solvent floating-zone method \cite{Motoyama}, while Sr$_2$CuO$_2$Cl$_2$ is 
grown by cooling the stoichiometric melt \cite{Perkins}. Thin flakes with 
thickness of 50-100 $\mu$m are cleaved out for transmission measurements. 
Two optical parametric generators pumped by a KHz regenerative amplifier 
generate pump and probe pulses. By using second harmonic and difference 
frequency generation techniques, the laser system can generate pulses with 
a temporal width of 0.2 ps and photon energy from 0.3 to 1.5 eV. We measure
differential transmission $\Delta T/T$, where $T$ is the transmission in 
the absence of the pump pulse, as a function of the pump-probe delay. The pump
and probe beams are polarized along the Cu-O chains (b axis in Sr$_2$CuO$_3$, 
c axis in SrCuO$_2$ and a axis in Sr$_2$CuO$_2$Cl$_2$).

The temporal behavior of the photoinduced absorption change at 290 K, 
$\Delta \alpha L$ = --ln(1+$\Delta T/T$), where $L$ is sample thickness, are
shown in Fig.~2 for all the three materials with the pump intensity 
$I_{pump}\sim $ 0.2 GW/cm$^{2}$. The 2D material shows absorption change for
both co- and cross-polarizations of the pump and probe beams, regardless of
the angle between the polarization of the pump light and the crystallographic 
axes. We have found that $\Delta \alpha L$ $\propto I_{pump}$ up to 
$I_{pump}\sim $ 10 GW/cm$^{2}$ and $I_{pump}\sim $ 1 GW/cm$^{2}$ 
for 1D and 2D materials, respectively. This indicates that the nonlinear effect
is third-order in the light field up to these $I_{pump}$. One can observe 
from Fig.~2(a) that the temporal profile of $\Delta \alpha L$ in 
Sr$_{2}$CuO$_{3}$ consists of a prompt component, which is determined by the 
laser pulse duration, and a slowly decaying component with characteristic time 
$\sim $ 1 ps. From comparisons of the temporal profiles at various pump photon 
energies we conclude that the relative magnitude of the decay component 
decreases with decreasing pump photon energy. The temporal profile of 
$\Delta \alpha L$ in SrCuO$_{2}$ (Fig.~2(b)) is similar to that in 
Sr$_{2}$CuO$_{3}$. On the other hand, in Sr$_{2}$CuO$_{2}$Cl$_{2}$, there 
exists a slower component with characteristic time $\sim $ 30 ps as well as the
faster component with $\sim $ 1 ps (see the upper panel of Fig.~2).

The prompt component of $\Delta \alpha L$ is due to the coherent optical
nonlinearity associated with TPA. In Fig.~3 we have shown the TPA spectra, 
plotted against $\omega_{pump} +\omega_{probe}$, for all three materials. In 
all cases, one- and two-photon absorption maxima nearly coincide indicating 
that overlapping TPA and linear absorption bands is a common feature of 
cuprates. The TPA spectrum of Sr$_2$CuO$_3$ at 290 K is similar to that 
obtained at 10 K, with maximum $\beta \sim$ 150 cm/GW at 2.1 eV 
\cite{Ogasawara}. We further observe from Fig.~3 that while in the 1D systems 
Sr$_2$CuO$_3$ and SrCuO$_2$ $\beta$ is of the order of 100 cm/GW and the width 
of the TPA band is about 0.5 eV, in Sr$_2$CuO$_2$Cl$_2$ the TPA coefficient is 
one order of magnitude smaller and the TPA band is broader. In particular, the 
low energy tail of the TPA in Sr$_2$CuO$_2$Cl$_2$ continues down to 
${\omega_{pump} + \omega_{probe} \sim}$ 1.5 eV, while the TPA coefficient in 1D
materials falls below 0.1 cm/GW at 1.5 eV.

In order to understand the differences in the TPA spectra between the 1D and 2D
systems, we adopted the two-band extended Hubbard model, which enables us to 
take into account the CT nature of the excited states explicitly, 
\begin{eqnarray}
H &=&-t\sum_{<ij,\sigma }(c_{i\sigma }^{\dagger }c_{j\sigma }+c_{j\sigma
}c_{i\sigma }^{\dagger })+\sum_{i}U_{i}n_{i\uparrow }n_{i\downarrow } 
\nonumber \\
&+&V\sum_{<ij>}n_{i}n_{j}+\sum_{i}\epsilon _{i}n_{i},  \label{hamiltonian}
\end{eqnarray}
where $c_{i\sigma }^{\dagger }$ creates a hole with spin $\sigma $ on site 
$i$, $n_{i\sigma }=c_{i\sigma }^{\dagger }c_{i\sigma }$, $n_{i}=\sum_{\sigma
}n_{i\sigma }$, and $<ij>$ implies nearest-neighbor sites. In Eq.~
(\ref{hamiltonian}), $t$ is the hopping between Cu and O sites, $U_{i}$ is 
the on-site Coulomb repulsion between two holes (different on Cu and O sites), 
$V$ is the Coulomb repulsion between holes on neighboring Cu and O, and 
$\epsilon _{O}-\epsilon _{Cu}$ is the site energy difference between O and Cu 
sites.

It is instructive to first consider the qualitative difference between 1D
and 2D within Eq.~(\ref{hamiltonian}). There exists a single mirror plane 
in 1D, and optical transitions are between symmetry subspaces that are 
``plus'' and ``minus'' with respect to this mirror plane. Consider now a 
three-atom segment OCuO. The single hole in the segment can occupy the Cu-site 
or either of the two O-sites. We denote these by the ``cartoon'' configurations
010, and 100 and 001, respectively. The ground state, as well as the excited + 
symmetry two-photon state are superpositions of 010 and (100 + 001), while the 
-- symmetry optical state is simply (100 -- 001). The dipole operator $\mu $ 
within Eq.~(\ref{hamiltonian}) is $\sum_{i}\vec{r_{i}}n_{i}$ (we take 
electronic charge $e$ = 1), where $\vec{r_{i}}$ gives the vector location of 
each atom, and has nonzero matrix element between the configurations 
(100 + 001) and (100 --- 001). The strength of the dipole coupling between 
the one-photon state and the ground (two-photon) state, 
$\mu _{01}$ ($\mu _{12}$), then depends on the ``overlap'' between the 
(100 -- 001) one-photon state and the (100 + 001) component of the ground 
(two-photon) state.

Consider now the 2D case, where each plaquette consists of one Cu and four 
O-atoms, again with a single hole. There occur two mirror planes now, and the 
symmetry subspaces are + +, + --, -- + and -- -- with respect to these. 
Eigenstates in the + + subspace are now superpositions of the configuration 
with the hole on the Cu, and {\it four} configurations with the hole 
occupying the different O-atoms. This subspace contains both the ground state 
and a two-photon state. The optical states are in the +-- and --+ subspaces, 
are degenerate, and are still superpositions of only {\it two} configurations 
each: the +-- eigenstate is a superposition of the two configurations with 
holes on the O-atoms to the left and to the right of the central Cu, while the 
--+ eigenstate is composed of the configurations with the holes above and below
the Cu. The dipole coupling between the one-photon and ++ states then can 
involve only two of the four configurations with holes on O-sites. Therefore, 
the transition dipole moments between the ground and one-photon states, 
$\mu_{01}$, and one- and two-photon states, $\mu _{12}$, are smaller in the 2D 
than in 1D. Since $\alpha $ and $\beta $ are proportional to $\mu_{01}^{2}$ 
and $\mu _{01}^{2}\mu _{12}^{2}$, respectively \cite{Boyd}, we expect both 
linear absorption and TPA to be weaker in 2D than in 1D, with the reduction in 
TPA strength larger. The above physical arguments for small-size cluster 
picture are valid even with a single electron, indicating that the weaker TPA 
in 2D can be explained without introducing charge-spin decoupling in 1D, as 
has been previously claimed \cite{Mizuno}.

We have verified the above qualitative reasonings by exact diagonalization 
studies of 1D and 2D clusters containing 4 Cu atoms (see Fig.~\ref{theory}).
The 2D cluster chosen is the largest system that can be diagonalized exactly. 
The choice of the 1D cluster size was based on the requirement that the number 
of holes are the same in 1D and 2D. The parameters considered were $|t|$ = 1.4 
eV, $U_{{\rm {Cu}}}$ = 10 eV, $U_{{\rm {O}}}$ = 3 eV, $V$ = 1 eV and 
$\epsilon_{{\rm {O}}}-\epsilon_{{\rm {Cu}}}$ = 2 eV. In Fig.~\ref{theory}, 
we have shown the lowest energy levels, along with the mirror planes $\sigma$ 
in 1D and $\sigma_x$ and $\sigma_y$ in 2D. The very low energy excited states 
in the + subspace in 1D and + + subspace in 2D in Fig.~4 are spin excitations 
that play no direct role in optical nonlinearity and are not discussed further.
The calculated electronic structure and the identification of optically 
relevant states are consistent with the recent group-theoretical analysis 
based on the excitonic cluster model \cite{Schulzgen}.

Our numerical results can be summarized as follows:
(i) The dipole couplings of the ground state with the lowest pair of + -- and 
-- + states in 2D are one order of magnitude smaller than that with the next 
higher pair of + -- and -- + states in the 2D lattice. Accordingly, the 
optical states in 2D is a {\it higher energy} pair of + -- and -- + states 
(see Fig.~4). $\mu_{01}^{2}$\ in 2D is 0.36 in our units, as compared to 
$\mu _{01}^{2}\simeq 1$ in the 1D cluster, thereby explaining the weaker 
linear absorption in 2D (see Fig.~3). 
(ii) As seen in Fig.~\ref{theory}, in 2D, there exist even parity states,
which are antisymmetric with respect to both $\sigma_{x}$ and $\sigma_{y}$ 
(-- -- subspace). TPA to these states should occur for cross-polarized pump 
and probe beams. We have observed the TPA with the cross-polarized 
configuration in our experiments and will discuss these elsewhere. Note that 
any interaction (electron-phonon interaction, intrinsic asymmetry due to 
crystal twinning) that leads to violation of this strict symmetry principle 
will cause weak TPA to -- -- states even with co-polarized pump and probe. It 
is conceivable that the broad nature of the TPA in 2D in the high energy 
region and the persistence of weak TPA in the low energy region ($\sim $ 1.5 eV)in Sr$_2$CuO$_2$Cl$_2$ (see Fig.~3) are due to the -- -- states. As seen in 
Fig.~\ref{theory}, there also exists a -- -- state that occurs below the true 
optical states. This state was not found in the previous one-band model 
\cite{Mizuno} (presumably because the smaller lattice investigated there did 
not have the symmetry of the 2D lattice). 
(iii) In 2D, the most dominant two-photon states are the lowest and the 
third charge excitation states in the + + subspace, whose dipole couplings are 
1.25 and 1.35 to the two one-photon states, compared to dipole couplings of 
3.4 and 1.2 between the dominant two-photon states and the optical state in 1D.
The lower of the two + + excited states occurs slightly below the one-photon 
states in 2D for the parameters we have chosen, but with small modifications of
these parameters it can also occur slightly above the one-photon states. Thus 
in both 1D and 2D the dominant two-photon states are close to the one-photon 
states in energy, in agreement with experiment. The numerical simulation 
predicts TPA in 1D larger by roughly one order of magnitude than in 2D even 
from these small cluster calculations. As discussed above, this is a 
consequence of the larger coordination number in 2D. 
(iv) We have included direct O--O hopping up to $|t|$/3 into Hamiltonian 
(\ref{hamiltonian}) and found no changes in our above conclusions.

The other distinct dimensionality dependence is the difference in the 
relaxation time. Specifically, in Sr$_2$CuO$_2$Cl$_2$, there exist decay 
components with characteristic time $\sim $ 30 ps and $\sim $ 1 ps, while 1D 
materials show only the latter component. The intensity of the decay 
components reduces with an decrease in pump photon energy (see Fig.~2). This 
indicates that the decay mechanism is associated with excitation of real 
carriers. We ascribe the 1 ps relaxation to non-radiative channels through 
spinon (in 1D) or magnon (in 2D) states, which exist below the optical gap and 
whose width is $\sim$ 1 eV \cite{Matsuda,Ogasawara}. We ascribe the slower 30 
ps optical relaxation in Sr$_2$CuO$_2$Cl$_2$ to the existence of low energy + 
-- and -- + states with weak dipole coupling to the ground state and the 
forbidden -- -- electronic state that can act as trap states upon 
photoexcitation (see Fig.~4). Such subgap electronic excitations are absent in 
1D, and therefore the trapping of the optical excitation cannot occur. We note 
that the undoped 2D cuprates YBa$_2$Cu$_3$O$_6$ and Nd$_2$CuO$_4$ have also 
shown two decay components with comparable decay rates \cite{Matsuda}.

In conclusion, pronounced dimensionality dependence of the optical nonlinearity
in the cuprates is found experimentally and theoretically. Coherent optical 
nonlinearity is dominated by two-photon states whose energy locations relative 
to the optical states are same in both 1D and 2D materials. However, large 
nonlinearity and ultrafast relaxation are characteristics of only 1D SCE 
systems, which make them promising materials for ultrafast optoelectronics 
\cite{Ogasawara}. The spin excitations with large energies, promoting the 
relaxation, are unique to SCE systems. The smaller nonlinearity in 2D is due 
to the larger coordination number of the atoms, and the occurrence of a slower 
relaxation channel in 2D is most probably associated with a fundamental 
difference in the electronic structures in 1D and 2D, viz., the occurrence of 
electronic energy states below the optical states in 2D, and their absence in 
1D.

We are very grateful to T. Ogasawara and S. Uchida for their cooperation at 
the early stage of this work. We also thank M. Nagai and R. Shimano for 
stimulating discussions. This work was supported by a grant-in-aid for COE 
Research from the Ministry of Education, Science, Sports, and Culture of Japan 
and the New Energy and Industrial Technology Development Organization (NEDO). 
Work in Arizona was supported by the U.S. NSF.

\begin{figure}[tbp]
\caption{Crystal structures of cuprates. (a) Sr$_2$CuO$_3$ (b=3.91 A). (b)
SrCuO$_2$ (c=3.92 A). (c) Sr$_2$CuO$_2$Cl$_2$ (a= b=3.98 A). Schematics of
the Cu-O networks are also shown.}
\end{figure}

\begin{figure}[tbp]
\caption{Temporal profiles of $\Delta \protect\alpha L$, which are normalized 
at their maximum values, at 290K in Sr$_2$CuO$_3$ (a), SrCuO$_2$ (b), and 
Sr$_2$CuO$_2$Cl$_2$ (c) for (pump + probe) energies of (1.46 + 0.95 eV) in the 
upper part and (0.95 + 1.03eV) in the lower part. One can observe that the 
picosecond decay components are pronounced for the higher pump photon energy. 
The upper panel shows $\Delta \protect\alpha L$, now plotted on a logarithmic 
scale, for the upper cases ((a), blue; (b), green; (c), red) plotted against 
extended time scale. The black solid line shows a fitting curve to the 
$\Delta \protect\alpha L$ of Sr$_2$CuO$_2$Cl$_2$ with two exponential functions
with decay times of $\tau$=1 ps and $\tau$=30 ps.}
\end{figure}

\begin{figure}[tbp]
\caption{TPA coefficient $\protect\beta$ versus 
$\protect\omega_{pump} + \protect\omega_{probe}$ at 290K in Sr$_2$CuO$_3$ (a), 
SrCuO$_2$ (b) and Sr$_2$CuO$_2$Cl$_2$ (c). Circles, squares and triangles 
correspond to pump energies at 1.46, 1.31 and 0.95 eV, respectively. Solid 
lines indicate the linear absorption $\protect\alpha$.}
\end{figure}

\begin{figure}[tbp]
\caption{Cluster models studied numerically and the corresponding energy
levels in 1D (a) and 2D (b). Filled (open) circles are Cu (O) sites. Dashed 
lines denote mirror-plane symmetries used. States labeled * (1-photon states) 
have large dipole coupling to the ground state, those labeled \# (2-photon
states) have large dipole coupling to 1-photon states. See text for 
parameters.}
\label{theory}
\end{figure}

\end{document}